\documentclass[pdflatex,sn-mathphys-num]{sn-jnl}

\usepackage{graphicx}%
\usepackage{multirow}%
\usepackage{amsmath,amssymb,amsfonts}%
\usepackage{amsthm}%
\usepackage{mathrsfs}%
\usepackage[title]{appendix}%
\usepackage{xcolor}%
\usepackage{textcomp}%
\usepackage{manyfoot}%
\usepackage{booktabs}%
\usepackage{algorithm}%
\usepackage{algorithmicx}%
\usepackage{algpseudocode}%
\usepackage{listings}%
\usepackage{amsmath}
\usepackage{algpseudocode}
\usepackage{algorithm}
\usepackage{geometry}
\usepackage[T1]{fontenc}
\usepackage{lmodern}

\theoremstyle{thmstyleone}%
\theoremstyle{thmstyletwo}%

\theoremstyle{thmstylethree}%

\begin{document}

\title[Article Title]{A Cross-Chain Event-Driven Data Infrastructure for Aave Protocol Analytics and Applications}


\author[1]{\fnm{Junyi} \sur{Fan}}\email{junyifan@usc.edu}
\author[1]{\fnm{Li} \sur{Sun}}\email{lsun4765@usc.edu}


\affil*[1]{\orgdiv{Daniel J. Epstein Department of Industrial \& Systems Engineering}, \orgname{University of Southern California}, \orgaddress{\street{3715 McClintock Ave GER 240}, \city{Los Angeles}, \postcode{90007}, \state{California}, \country{United States}}}


\abstract
{Decentralized lending protocols, exemplified by Aave V3, have transformed financial intermediation by enabling permissionless, multi-chain borrowing and lending without intermediaries. Despite managing over \$10 billion in total value locked, empirical research remains severely constrained by the lack of standardized, cross-chain event-level datasets. 
This paper introduces the first comprehensive, event-driven data infrastructure for Aave V3 spanning six major EVM-compatible chains (Ethereum, Arbitrum, Optimism, Polygon, Avalanche, and Base) from respective deployment blocks through October 2025. We collect and fully decode eight core event types — Supply, Borrow, Withdraw, Repay, LiquidationCall, FlashLoan, ReserveDataUpdated, and MintedToTreasury — producing over 50 million structured records enriched with block metadata and USD valuations. 
Using an open-source Python pipeline with dynamic batch sizing and automatic sharding (each file $\leq$ 1 million rows), we ensure strict chronological ordering and full reproducibility. The resulting publicly available dataset enables granular analysis of capital flows, interest rate dynamics, liquidation cascades, and cross-chain user behavior, providing a foundational resource for future studies on decentralized lending markets and systemic risk.}

\keywords{decentralized finance, blockchain data, event-driven analytics}



\maketitle

\section{Background and Summary}

The global financial system, with lending markets exceeding \$150 trillion annually\cite{ref1}, has long relied on centralized intermediaries that introduce inefficiencies, accessibility barriers, and systemic risks. Blockchain technology emerged as a transformative solution, enabling decentralized finance (DeFi) protocols that provide financial services without traditional intermediaries\cite{ref2}. The rapid growth of DeFi, with total value locked (TVL) surpassing \$200 billion at its peak, represents a fundamental shift toward programmable, transparent financial infrastructure\cite{ref3}.

Among DeFi protocols, lending platforms have emerged as critical infrastructure addressing fundamental inefficiencies in traditional financial intermediation. Aave exemplifies this transformation by creating a comprehensive lending ecosystem that solves multiple challenges inherent in both traditional finance and early DeFi implementations\cite{ref4}. The protocol addresses three core problems: capital inefficiency in traditional banking systems, lack of programmable financial primitives, and limited access to sophisticated financial instruments for retail users\cite{ref5}.

Aave's approach to decentralized lending centers on liquidity pools that aggregate user deposits, enabling instant borrowing without requiring direct lender-borrower matching\cite{ref6}. This pool-based model eliminates the matching inefficiencies of peer-to-peer systems while providing continuous liquidity for both suppliers seeking yield and borrowers requiring capital. The protocol's interest rate mechanisms automatically adjust based on supply and demand dynamics, creating market-driven pricing that responds to real-time liquidity conditions without manual intervention\cite{ref7}. Suppliers deposit assets into these pools and receive interest-bearing aTokens that continuously accrue value, while borrowers can access instant liquidity by providing over-collateralized positions that protect the protocol from default risk\cite{ref8}.

Beyond basic lending functionality, Aave introduces innovative financial primitives that extend traditional banking concepts into programmable, composable forms. Flash loans enable uncollateralized borrowing within single transactions, creating arbitrage opportunities and capital efficiency mechanisms impossible in traditional finance\cite{ref9}. Credit delegation allows users to delegate borrowing power to other addresses, enabling sophisticated use cases such as institutional borrowing and yield farming strategies\cite{ref10}. The protocol's collateral management system supports multiple asset types with dynamic risk parameters, allowing users to optimize their capital efficiency through features like collateral swapping and debt position management\cite{ref11}. The protocol's multi-chain deployment strategy addresses scalability and accessibility challenges by providing lending services across multiple blockchain networks with varying cost structures and performance characteristics\cite{ref12}. Advanced features like cross-chain asset portals and efficiency modes for correlated assets further enhance capital utilization while maintaining robust risk management frameworks\cite{ref13}. Aave V3's current ecosystem manages over \$5 billion in total value locked across all deployments, with the protocol demonstrating remarkable resilience through multiple market cycles including the Terra Luna collapse, FTX bankruptcy, and various DeFi exploits\cite{ref15}.

This comprehensive approach to decentralized lending creates rich behavioral data spanning multiple dimensions: user interactions across different functional modules, cross-chain activity patterns, risk management decisions, and market response mechanisms. The resulting event streams capture not just transaction volumes but the complete spectrum of user intentions, protocol state changes, and market dynamics that drive the DeFi ecosystem\cite{ref15}. Despite the transparent nature of blockchain data, DeFi research faces critical limitations that hinder comprehensive protocol analysis.

The lack of standardized, comprehensive datasets for major lending protocols has become a significant bottleneck for empirical research\cite{ref6}. While blockchain data is publicly available, extracting, standardizing, and validating protocol-specific event data across multiple chains requires substantial technical infrastructure that most researchers lack\cite{ref7}. Existing datasets are typically fragmented, covering single chains or limited time periods, making cross-chain behavioral analysis virtually impossible. Current analytical approaches remain superficial, focusing primarily on transaction volumes and basic statistical measures while missing the rich behavioral patterns inherent in protocol-specific interactions. Recent studies, such as the comprehensive Uniswap analysis by Chemaya et al.\cite{ref8}, provide valuable insights into transaction-level patterns and decentralization metrics but fall short of capturing the complex risk dynamics, user behavioral sequences, and cross-protocol interactions that drive DeFi ecosystem evolution\cite{ref9}.

The multi-chain nature of modern DeFi protocols compounds these analytical challenges. Users increasingly operate across multiple networks, creating behavioral patterns that span different blockchain environments\cite{ref10}. Understanding these cross-chain dynamics requires unified analytical frameworks capable of processing heterogeneous event streams while preserving the semantic richness of protocol-specific interactions\cite{ref11}. Current methodologies lack the sophistication to model these complex, multi-dimensional relationships, limiting our understanding of systemic risks and user behavior evolution in the expanding DeFi landscape\cite{ref12}.

To address these critical gaps in DeFi research, this paper introduces a standardized, event-driven dataset for comprehensive analysis of the Aave protocol across six major blockchain networks. Our dataset captures the full spectrum of protocol interactions, enabling more thorough and flexible analysis than traditional transaction-level data alone. We make two primary contributions that advance the field of DeFi analytics. First, we present the first standardized dataset spanning six major blockchain networks, capturing eight distinct event types that cover everything from core financial operations to advanced protocol functionalities and critical risk management events. Each event is enriched with detailed metadata, including gas costs, block information, timestamps, and standardized USD valuations, providing researchers with unprecedented cross-chain analytical capabilities. Second, we provide a complete open-source implementation of our data extraction and analysis pipeline, ensuring full reproducibility and transparency for the research community. This infrastructure enables researchers to perform cross-chain blockchain studies, analyze market sentiment, identify protocol and systemic risks, and explore community behavior patterns using descriptive statistics, econometric modeling, and advanced machine learning techniques.

The significance of these contributions extends beyond technical innovation to address critical gaps in DeFi research methodology. Our work establishes a new approach for DeFi analytics that bridges the gap between raw blockchain data and actionable financial insights, enabling investigation of complex phenomena such as systemic risk propagation, cross-chain user behavior evolution, and protocol optimization strategies that were previously inaccessible to researchers.

\section{Main Logic}
\subsection{Supply Logic}
\begin{figure}[h]
    \centering
    \includegraphics[width=1.0\textwidth]{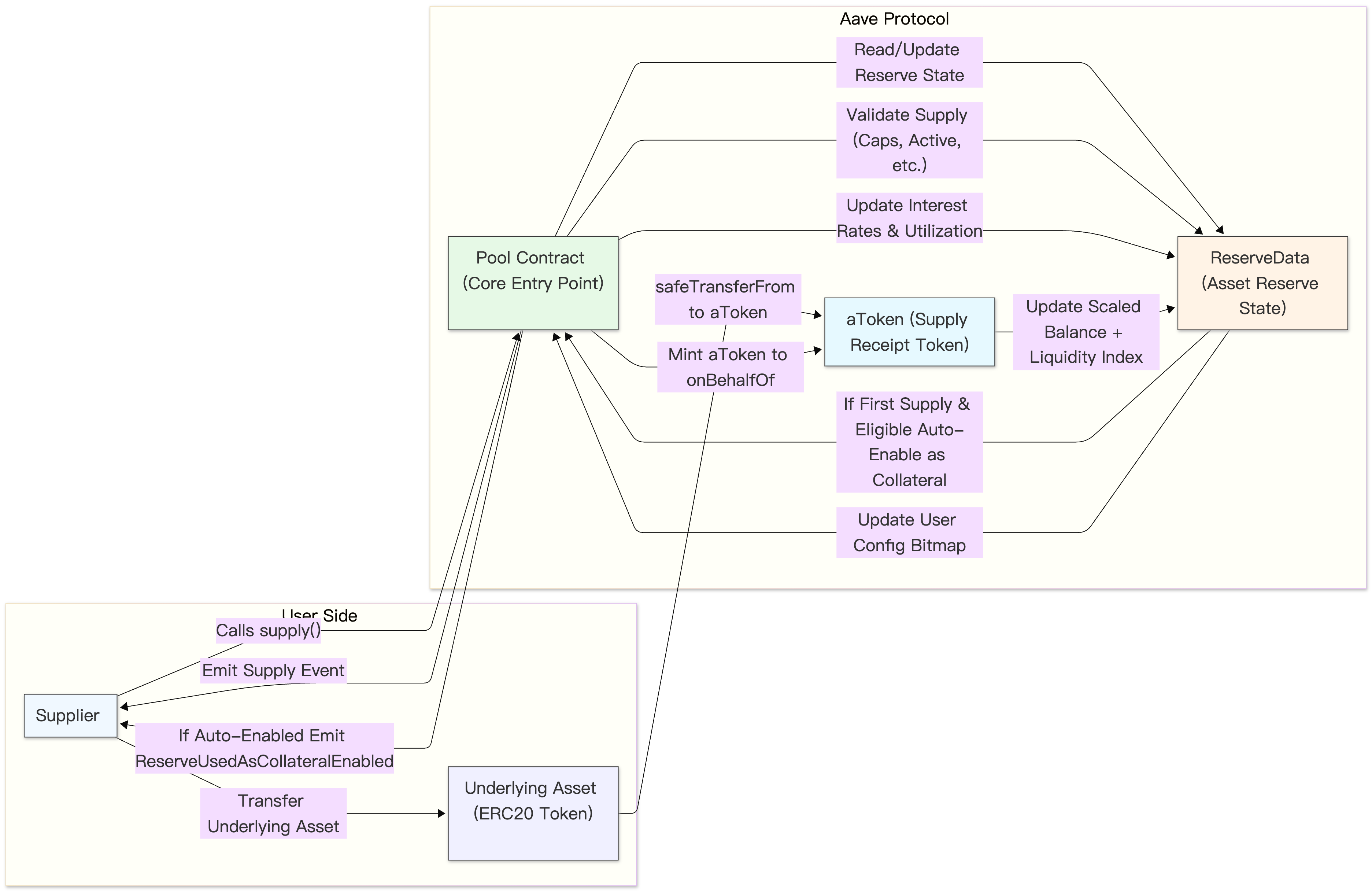}
    \caption{Supply logic of Aave V3 protocal.}
    \label{fig:supply_logic}
\end{figure}

\begin{figure}[h]
    \centering
    \includegraphics[width=1.0\textwidth]{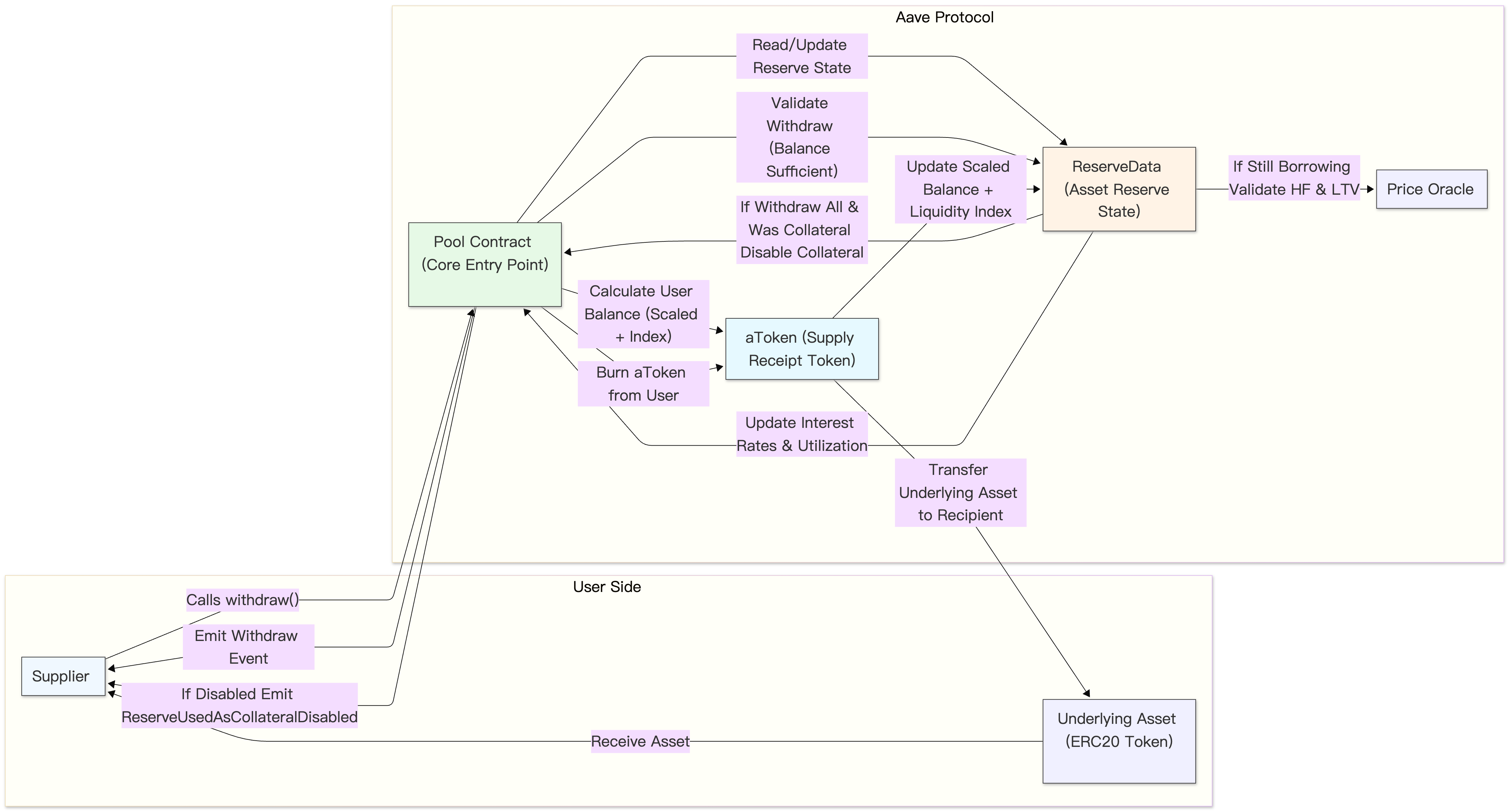}
    \caption{Withdraw logic of Aave V3 protocal.}
    \label{fig:withdraw_logic}
\end{figure}

The supply mechanism in Aave V3 is like depositing money into a savings account that also serves as collateral for future loans. Users transfer assets into the protocol and instantly receive interest-bearing tokens (aTokens) in return. These tokens grow in value over time as interest accrues, and the deposited assets can be used as backing to borrow other currencies without selling anything.
Unlike a traditional bank where your deposit just sits and earns a fixed rate, here the money is actively lent out to borrowers, and you earn a market-driven yield. You can withdraw anytime, as long as it doesn’t jeopardize your borrowing capacity. If you have loans open, the system checks that enough collateral remains before allowing full withdrawal—just like a bank ensuring you don’t drain your account while a mortgage is active.
Users can also choose whether to use their deposit as collateral. Turning it on increases borrowing power; turning it off gives more withdrawal freedom. The system automatically enables collateral for first-time deposits if the asset qualifies, similar to a bank auto-activating overdraft protection.
Transfers of aTokens between users are monitored: if someone gives away all their position in an asset used as collateral, the system disables it for them and may activate it for the recipient—ensuring the protocol always knows who is responsible for maintaining loan health.
The full execution flow is shown in Figure~\ref{fig:supply_logic} and Figure~\ref{fig:withdraw_logic}.
This creates a fluid, efficient savings-and-loan system: deposit once, earn yield, borrow against it, withdraw when needed—all instantly, globally, and without intermediaries.


\subsection{Borrow Logic}

\begin{figure}[h]
    \centering
    \includegraphics[width=1.0\textwidth]{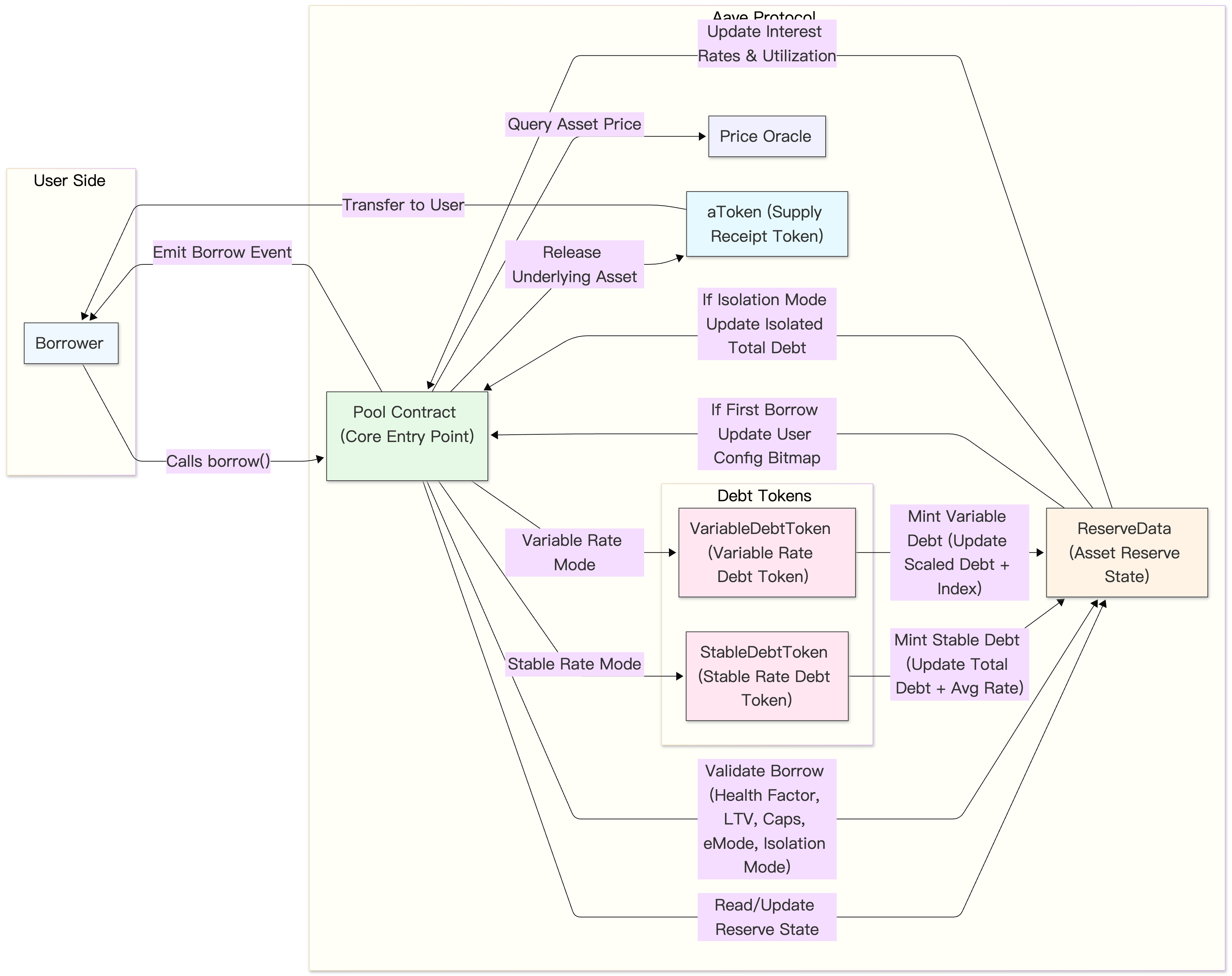}
    \caption{Borrow logic of Aave V3 protocal.}
    \label{fig:borrow_logic}
\end{figure}

\begin{figure}[h]
    \centering
    \includegraphics[width=1.0\textwidth]{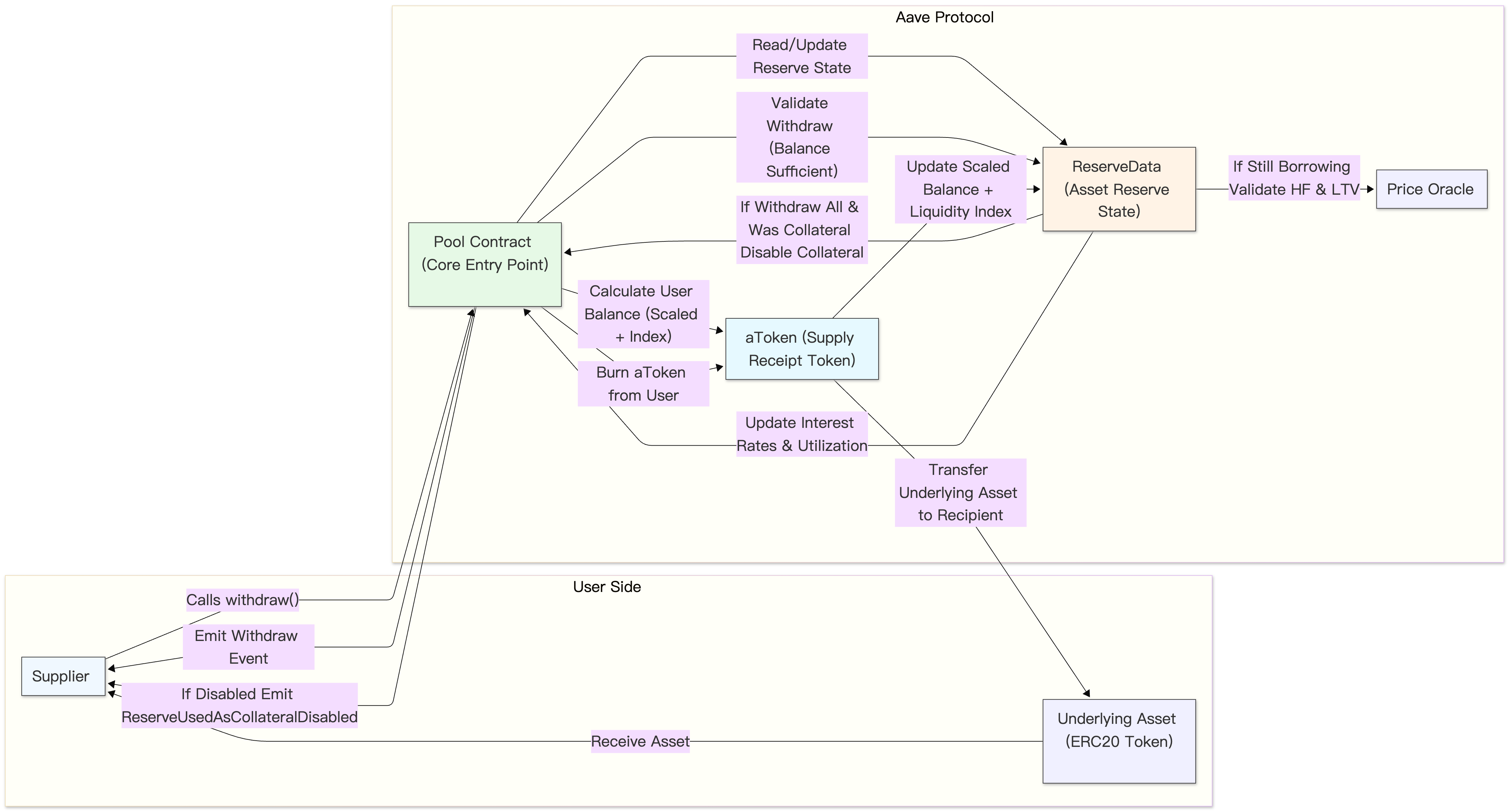}
    \caption{Repay logic of Aave V3 protocal.}
    \label{fig:repay_logic}
\end{figure}

The borrow mechanism in Aave V3 functions like a collateralized line of credit. Users deposit assets as collateral and can immediately borrow other assets without selling their holdings. This allows them to maintain their positions while accessing liquidity for trading, investing, or daily needs.
In traditional finance, loans require credit checks, paperwork, and days of processing. In Aave, the process is instant and automated through smart contracts—no banks, no delays, fully transparent on-chain.
The flow is straightforward. When borrowing, the system evaluates the user’s collateral to confirm it supports the loan amount. If approved, a debt position is created, interest begins accruing, and the requested asset is transferred to the user’s wallet. Rates adjust dynamically based on market utilization.
Repayment works in reverse: return the borrowed asset plus interest, and the system reduces the debt. Once fully repaid, borrowing status is cleared, and collateral is fully usable again.
Users can also switch between stable and variable interest rates during the loan, similar to refinancing. For higher-risk assets, Aave uses isolation mode to cap total borrowing and reduce systemic risk.
The complete execution steps are detailed in Figure~\ref{fig:borrow_logic} and Figure~\ref{fig:repay_logic}.
This system enables true capital efficiency in DeFi: leverage what you own to access what you need, all while earning yield on deposits and paying market-driven rates—accessible globally with just a crypto wallet.


\subsection{Reserve Logic}
The ReserveLogic in Aave V3 is like the central ledger of a cooperative bank that continuously tracks every penny of deposits, loans, and interest across all members. Every time someone deposits, borrows, or repays, this ledger recalculates how much interest has accrued since the last update—for both savers (who earn yield via linear index growth) and variable-rate borrowers (whose debt grows through continuous compounding).

It works in real time: before processing any user action, the system first creates a lightweight in-memory snapshot of the current reserve state (current indices, scaled debt amounts, rates, and configuration) to eliminate redundant storage reads. It then advances the two core cumulative indices to the exact current block, accurately reflecting elapsed-time interest on both the supply and borrow sides. At the same moment, the freshly accrued total debt interest is measured, and the configured reserve-factor portion is automatically diverted to the protocol treasury as scaled claims—building a community-controlled risk and revenue buffer, exactly like a credit union retaining a percentage of all interest income.

Once the user action has shifted available liquidity or total debt, the updated utilization metrics are passed to the reserve-specific interest-rate strategy, which instantly recomputes the next supply yield and borrowing rates based on real-time supply–demand balance, just as a bank dynamically adjusts its posted rates. The new rates are persisted and publicly emitted, effectively posting the daily interest-rate sheet on-chain. The complete execution flow is shown in Table~\ref{tab:reservelogic}.

This tightly optimized, index-driven mechanism ensures every user’s balance is always up-to-date to the current block, yield distribution remains perfectly fair, debt accounting stays precise, treasury revenue accrues transparently, and the entire protocol remains solvent—automatically, verifiably, and without any human intervention.

\begin{table}[h]
\caption{Reserve Logic}
\label{tab:reservelogic}
\begin{tabular}{|l|l|}
\hline
\textbf{Description} & \textbf{Functions Triggered} \\
\hline
\multicolumn{2}{|l|}{\textbf{RESERVEDATAUPDATED}} \\
\hline
Take snapshot of current reserve state. & \texttt{reserve.cache()} \\
\hline
Skip update if already done this block. & \texttt{lastUpdateTimestamp == block.timestamp} \\
\hline
Calculate interest accrued for suppliers since last update. & \texttt{calculateLinearInterest(liquidityRate, timestamp)} \\
\hline
Update supplier yield index. & \texttt{nextLiquidityIndex =} \\ 
& \texttt{cumulated.rayMul(currLiquidityIndex)} \\
\hline
Calculate compounded interest for variable borrowers. & \texttt{calculateCompoundedInterest(variableRate, timestamp)} \\
\hline
Update borrower debt index. & \texttt{nextVariableBorrowIndex =} \\ 
 & \texttt{cumulated.rayMul(currVariableBorrowIndex)} \\
\hline
Compute total new debt created by interest. & \texttt{totalDebtAccrued = currentDebt - previousDebt} \\
\hline
Take protocol fee from new debt (reserve factor). & \texttt{percentMul(reserveFactor)} \\
\hline
Add fee to treasury (scaled by liquidity index). & \texttt{accruedToTreasury += amount.rayDiv(nextLiquidityIndex)} \\
\hline
Calculate new real-time interest rates. & \texttt{calculateInterestRates(...)} \\
\hline
Store updated rates (liquidity, stable, variable). & \texttt{currentLiquidityRate = ...} \\
\hline
Update timestamp to current block. & \texttt{lastUpdateTimestamp = block.timestamp} \\
\hline
Broadcast new rates and indices. & \texttt{emit ReserveDataUpdated(...)} \\
\hline
\end{tabular}
\end{table}

\subsection{Liquidation Logic}

The liquidation process in Aave V3 is executed by the \texttt{LiquidationLogic} library whenever a user's Health Factor drops below 1. The Health Factor is defined as
\begin{equation}
H = \frac{\sum_i C_i \, p_i \, \lambda_i}{\sum_j B_j \, p_j},
\end{equation}
where $C_i$ and $B_j$ are the collateral and borrow amounts of each asset (normalized to their respective decimals), $p_i$, $p_j$ are the corresponding USD oracle prices, and $\lambda_i \in [0,1]$ denotes the liquidation threshold of the $i$-th reserve.

A position is liquidatable if $H < 1$. The close factor $\kappa$ is
\begin{equation}
\kappa =
\begin{cases}
0.5  & \text{if } 0.95 < H < 1, \\
1.0  & \text{if } H \leq 0.95.
\end{cases}
\end{equation}
The actual debt repaid in the transaction is therefore $D = \min(\textit{debtToCover}, \kappa \cdot B)$.

Given $D$, the amount of collateral transferred to the liquidator is computed by the on-chain formula
\begin{align}
Q_{\text{base}} &= D \cdot \frac{p_D}{p_C} \cdot 10^{d_C - d_D}, \\[6pt]
Q_{\text{total}} &= Q_{\text{base}} \cdot (1 + \beta), \\[6pt]
F &= Q_{\text{total}} \cdot \beta \cdot \phi, \\[6pt]
Q_{\text{liq}} &= Q_{\text{total}} - F,
\end{align}
where
\begin{itemize}
  \item $p_D$, $p_C$: oracle prices of debt and collateral assets,
  \item $d_C$, $d_D$: decimals of collateral and debt assets,
  \item $\beta = \frac{LB - 10000}{10000}$: liquidation bonus rate (e.g., $LB = 10500 \Rightarrow \beta = 0.05$),
  \item $\phi \in [0,1]$: protocol fee share of the bonus (typically 0.2--0.3).
\end{itemize}

By conservatively calibrating the liquidation threshold $\lambda_i$ strictly below the loan-to-value ratio $\text{LTV}_i$ and maintaining a liquidation bonus rate of $\beta \approx 0.05$, the protocol enforces robust over-collateralization while ensuring strong economic incentives for timely liquidations, resulting in near-zero bad debt even during extreme market stress.

Thus, the liquidator receives $Q_{\text{liq}}$ units of collateral by repaying only $D$ units of debt, yielding an instantaneous risk-free profit of $F + Q_{\text{base}} - D$ in USD value, while the protocol collects fee $F$. This deterministic pricing mechanism has sustained near-zero bad debt across all Aave V3 markets.

\section{Methods}

In this study, we developed a systematic approach to extract and analyze event data from the Aave V3 lending protocol deployed across multiple blockchain networks. The primary objective was to collect historical event logs related to user interactions and protocol updates, enabling downstream analysis of lending dynamics, risk assessment, and multi-chain performance. Data extraction was performed using a custom Python-based script leveraging blockchain RPC endpoints, with a focus on efficiency, scalability, and data integrity. The process involved connecting to blockchain nodes, querying event logs in batches, decoding events, and storing results in structured formats. Below, we detail the data sources, tools, and procedures employed.

\subsection{Data Sources and Configuration}

Event data were sourced from the Aave V3 Pool contracts on six major Ethereum-compatible blockchains: Ethereum, Arbitrum, Optimism, Polygon, Avalanche, and Base. Each chain's Pool contract address, start block (corresponding to the protocol's deployment), and maximum block limit (set to October 1, 2025, 18:44 UTC for temporal consistency) were predefined. The configurations are summarized in Table~\ref{tab:blockchain-configs}.

\begin{table}[h]
\centering
\caption{Blockchain Configurations for Aave V3 Data Extraction}
\label{tab:blockchain-configs}
\begin{tabular}{l l r r}
\toprule
\textbf{Chain} & \textbf{Pool Address} & \textbf{Start Block} & \textbf{Max Block} \\
\midrule
Ethereum  & \texttt{0x87870Bca3F3fD6335C3F4ce8392D69350B4fA4E2} & 16,291,127 & 23,615,633 \\
Optimism  & \texttt{0x794a61358D6845594F94dc1DB02A252b5b4814aD} & 4,365,693  & 142,662,943 \\
Arbitrum  & \texttt{0x794a61358D6845594F94dc1DB02A252b5b4814aD} & 7,740,000  & 391,361,693 \\
Polygon   & \texttt{0x794a61358D6845594F94dc1DB02A252b5b4814aD} & 25,825,996 & 77,909,957 \\
Avalanche & \texttt{0x794a61358D6845594F94dc1DB02A252b5b4814aD} & 11,970,000 & 70,593,220 \\
Base      & \texttt{0xA238Dd80C259a72e81d7e4664a9801593F98d1c5} & 2,357,200  & 37,067,658 \\
\bottomrule
\end{tabular}
\end{table}

RPC connections were established via Alchemy API endpoints for reliable and high-throughput access. The script targeted 15 core event types emitted by the Pool contracts, including user actions (e.g., Supply, Borrow) and protocol updates (e.g., ReserveDataUpdated). Event signatures were mapped to human-readable names and decoding functions for structured extraction.

Table~\ref{tab:event_chain_matrix} presents the distribution of key Aave protocol events across six major blockchain networks. The events are ordered by their functional importance within the Aave ecosystem. ReserveDataUpdated maintains real-time pool state information, while Borrow, Repay, Supply, and Withdraw represent the core lending operations. LiquidationCall handles risk management through position liquidations, and the collateral events manage user collateral settings. The remaining events cover interest rate management, efficiency modes, isolation mechanisms, and protocol fee collection.

\begin{table}[h]
\centering
\caption{Chain Execution Count by Event Type}
\label{tab:event_chain_matrix}
{
\begin{tabular}{l|rrrrrr}
\toprule
\textbf{Event} & \textbf{Arbitrum} & \textbf{Avalanche} & \textbf{Base} & \textbf{Ethereum} & \textbf{Optimism} & \textbf{Polygon} \\
\midrule
ReserveDataUpdated               & 8,128,831 & 3,245,111 & 9,845,587 & 2,557,278 & 5,630,903 & 13,100,765\\
Borrow                           & 1,272,686 & 463,378   & 1,107,904 & 537,666   & 725,578   & 1,323,199 \\
Repay                            & 1,187,380 & 350,449   & 1,347,034 & 388,886   & 836,630   & 1,233,390 \\
Supply                           & 2,971,263 & 1,304,356 & 4,101,311 & 880,800   & 2,304,267 & 3,337,326 \\
Withdraw                         & 2,349,245 & 990,333   & 3,147,777 & 651,512   & 1,614,588 & 3,212,641 \\
LiquidationCall                  & 42,146    & 20,565    & 34,632    & 15,774    & 31,987    & 49,044    \\
ReserveUsedAsCollateralEnabled   & 2,044,860 & 532,006   & 2,596,129 & 598,376   & 1,084,462 & --        \\
ReserveUsedAsCollateralDisabled  & 1,758,619 & 438,129   & 2,139,644 & 497,820   & 885,422   & --        \\
RebalanceStableBorrowRate        & 1         & --        & --        & --        & --        & --        \\
UserEModeSet                     & 25,342    & --        & 10,853    & 21,637    & 41,468    & 39,287    \\
IsolationModeTotalDebtUpdated    & 18,240    & --        & 13        & 11,785    & 22,108    & 13,216    \\
MintedToTreasury                 & 8,382     & --        & 4,095     & 4,881     & 8,240     & 10,702    \\
\bottomrule
\end{tabular}
}
\end{table}

\subsection{Multi-Chain Extraction}

\begin{figure}[h]
    \centering
    \includegraphics[width=1.0\textwidth]{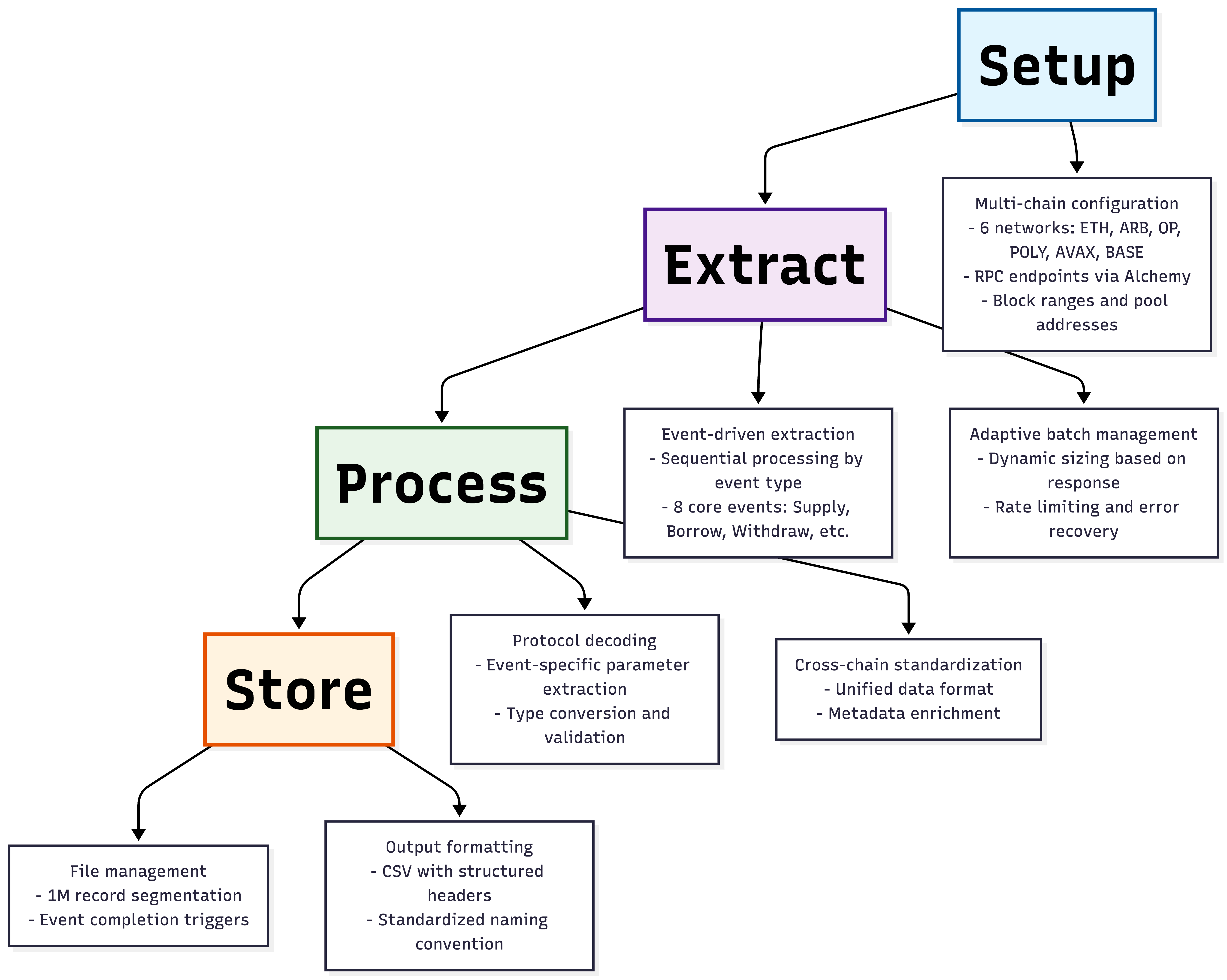}
    \caption{Basic .}
    \label{fig:extract_pipeline}
\end{figure}

Our data collection methodology employs an event-driven approach to systematically capture Aave protocol interactions across six major blockchain networks which is shown in Figure~\ref{fig:extract_pipeline}. The system implements a multi-chain architecture that standardizes data extraction while accommodating the unique characteristics of each blockchain environment. We focus on eight core event types that represent the fundamental operations within the Aave ecosystem, including basic lending operations such as supply, borrow, withdraw, and repay transactions, as well as advanced functionalities like flash loans and liquidation events.

The extraction process follows a sequential event-based strategy rather than chronological aggregation. For each blockchain network, we process one event type completely before moving to the next, ensuring data integrity and simplifying error recovery. This approach prevents event mixing complications while maintaining semantic coherence for each operation type. The system connects to blockchain networks through dedicated RPC endpoints, with each chain configured according to its specific parameters including pool contract addresses, deployment blocks, and network-specific constraints.

To handle the substantial volume of blockchain data efficiently, we implement dynamic batch sizing that automatically adjusts based on network response patterns. The system begins with large batch sizes for efficiency but intelligently reduces batch size when encountering rate limits or oversized responses. This adaptive mechanism ensures continuous data flow while respecting infrastructure constraints. The methodology incorporates comprehensive error handling that distinguishes between recoverable network issues and terminal errors, automatically retrying failed requests with adjusted parameters when appropriate.

Data organization follows a structured file management system that creates new files either when reaching one million records or when completing an event type. Each file follows a standardized naming convention that includes the blockchain network, event type, file sequence number, and timestamp. This segmentation strategy facilitates parallel processing, enables targeted analysis of specific event types, and provides natural breakpoints for quality assurance and validation procedures.

\section{Data Records}

The complete dataset is publicly available on Zenodo under DOI \href{https://doi.org/10.5281/zenodo.17898640}{10.5281/zenodo.17898640}.

It comprises fully decoded event-level data for Aave V3 across six EVM-compatible chains (Ethereum, Arbitrum, Optimism, Polygon, Avalanche, and Base), covering the period from each chain’s respective protocol deployment block up to the block heights corresponding to October 1, 2025 00:00 UTC. Eight core event types are included: Supply, Borrow, Withdraw, Repay, LiquidationCall, FlashLoan, ReserveDataUpdated, and MintedToTreasury.

Files are named as \texttt{aave\_V3\_\{chain\}\_\{event\}\_part\{nnn\}\_\{YYYYMMDD\_HHMMSS\}.csv}, where \texttt{\{chain\}} is the lowercase chain name, \texttt{\{event\}} is the CamelCase event name, \texttt{part\{nnn\}} is a zero-padded three-digit sequential part number that restarts independently for every (chain, event) pair, and \texttt{\{YYYYMMDD\_HHMMSS\}} denotes the exact timestamp at which the part file was written. 

Sharding is applied automatically per the moment the number of accumulated records for a specific (chain, event) combination reaches 1,000,000 or when scanning of that event is completed. Consequently, low-volume events are stored in a single \texttt{part001} file, whereas high-volume events (most notably ReserveDataUpdated on Ethereum) are split into multiple parts of at most one million rows each, preserving strict chronological order within each part. All fields are fully decoded and enriched with block and transaction metadata, making the files immediately usable without additional parsing. The open-source extraction and validation pipeline is included in the Zenodo record to ensure full reproducibility.

\section{Application}

\subsection{Data Aggregation Example}

\begin{figure}[h]
    \centering
    \includegraphics[width=0.8\textwidth]{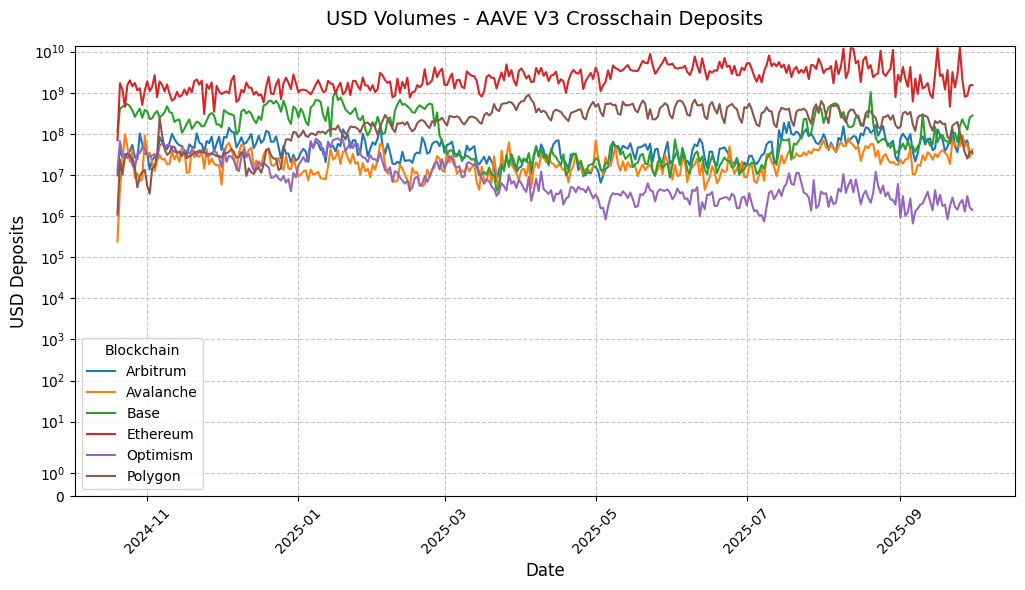}
    \caption{Cross-chain deposit volumes aggregated from our dataset.}
    \label{fig:crosschain_deposits}
\end{figure}

\begin{figure}[h]
    \centering
    \includegraphics[width=0.8\textwidth]{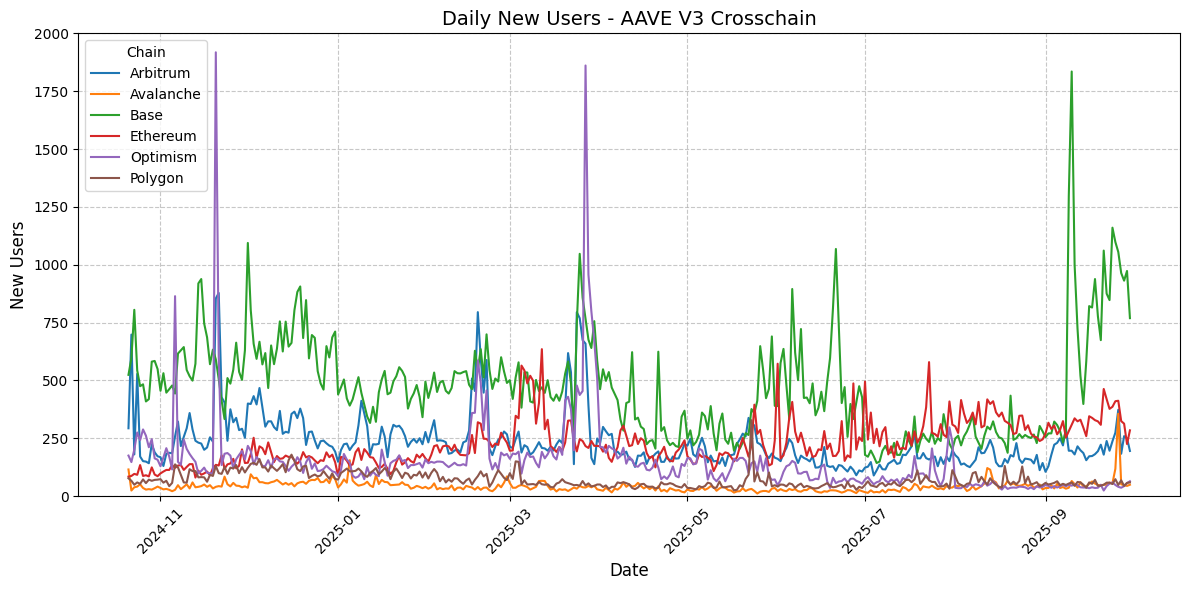}
    \caption{Daily new user trends derived from our dataset.}
    \label{fig:daily_new_users}
\end{figure}

\begin{figure}[h]
    \centering
    \includegraphics[width=0.8\textwidth]{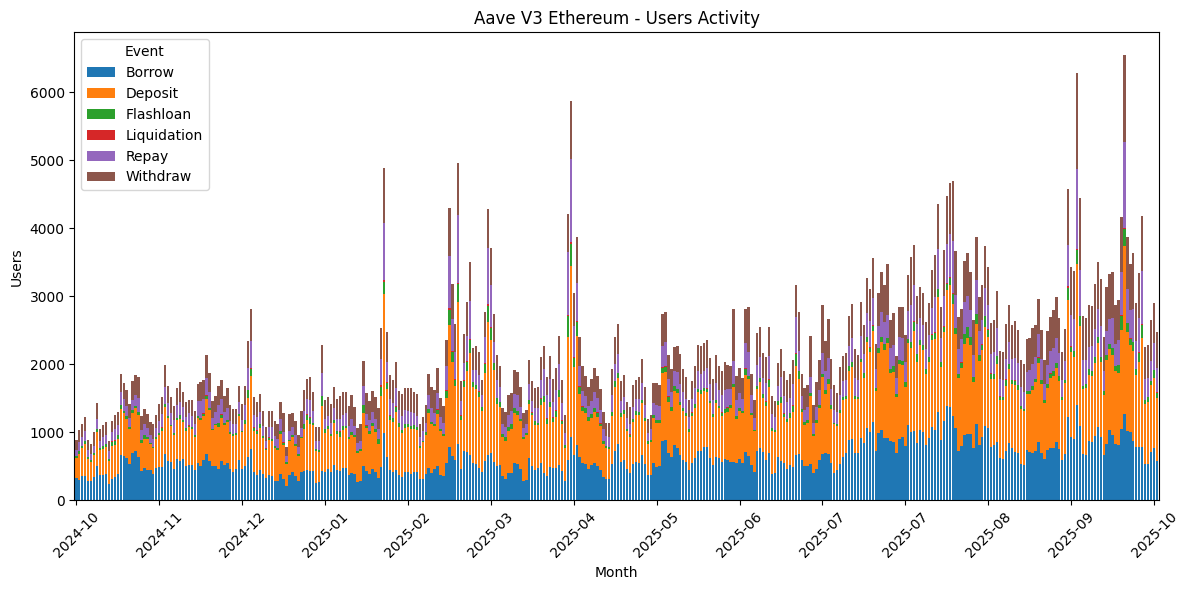}
    \caption{users activities count derived from our dataset.}
    \label{fig:users_activity}
\end{figure}

Our dataset supports robust descriptive statistical analysis by aggregating event data across chains and time periods. For instance, Figure~\ref{fig:crosschain_deposits} illustrates the total deposit volumes across multiple chains, highlighting variations in lending activity, with notable peaks on Arbitrum and Polygon likely due to higher transaction volumes or promotional events. Similarly, Figure~\ref{fig:daily_new_users} displays the daily influx of new users, revealing a steady increase with periodic spikes, possibly correlated with market trends or protocol updates. Figure~\ref{fig:users_activity} shows the activation times of various common events in the past year. It can be seen that the overall activation times of the Aave protocol have increased. The event-driven structure of our dataset enables seamless aggregation and visualization, accommodating large-scale datasets such as the millions of records captured for events like Supply and ReserveDataUpdated, as demonstrated by the flexibility in generating these statistical summaries.

\vspace{5.5cm}

\vspace{7.5cm}


\end{document}